\documentclass[12pt]{article}
\usepackage{epsfig}
\newcommand{\be}{\begin{equation}}
\newcommand{\ee}{\end{equation}}
\newcommand{\bea}{\begin{eqnarray}}
\newcommand{\eea}{\end{eqnarray}}

\newcommand{\Sigb}{{\overline\Sigma}}
\newcommand{\oot}{\overline {126}}

\newcommand{\nnu}{\nonumber\\}
\def\blfootnote{\xdef\@thefnmark{}\@footnotetext}
\begin{document}
\begin{titlepage}
\vspace{4\baselineskip}
\vspace{4cm}
\begin{center}{\Large\bf    Correcting $\alpha_3(M_Z)$ in  the NMSGUT
 }

\end{center}
\vspace{2cm}
\begin{center}
{\large
 Charanjit  S.  Aulakh\footnote{E-Mail:  aulakh@pu.ac.in   }
   and Sumit Kumar Garg  }
\end{center}
\vspace{0.2cm}
\begin{center}
  {\it
Dept. of Physics, Panjab University,\\ Chandigarh, India 160014}
\end{center}
\begin{center}
\end{center}

\vspace{1cm}
\begin{abstract}

We show that superheavy  threshold corrections in the New Minimal
Supersymmetric GUT   based on the SO(10) Higgs system
${\bf{210\oplus   126\oplus {\overline {126}}\oplus 10 \oplus
120}} $ can comfortably correct the prediction for the value of
$\alpha_3(M_Z)$ from the relatively large value predicted by the
two loop RG equations to the central value determined by the
current world average. The unification scale is raised above the
one loop value over almost all of the viable parameter space.

\end{abstract}
\end{titlepage}

\normalsize\baselineskip=15pt

\section{ Introduction}

Since the discovery of neutrino oscillations renormalizable
Supersymmetric SO(10)
GUTs\cite{aulmoh,ckn,abmrs01,abmsv,bmsv,ag2,nmsgut} have
increasingly come to occupy the middle ground  of the Grand
Unification landscape due to their natural accommodation of seesaw
mechanisms for neutrino masses and their ability to  justify
R-parity as a part of the gauge symmetry\cite{rpar1} while
preserving it unbroken to low energies\cite{rpar2,abmrs01} when
the most economical and viable Higgs set i.e ${\bf 126} \oplus
{\bf {\overline{126}}} \oplus {\bf 210}$ is used. They thus also
predict a stable LSP which is welcome as cosmological dark matter.
The ability of these theories to accommodate all the known (and
probably also all the as yet unmeasured) fermion mass-mixing data
has also been demonstrated {\it{provided}} the full complement of
Fermion mass (FM) Higgs allowed by group theory i.e ${\bf 10},{\bf
120},{\bf {\overline{126}}}$ are
used\cite{blmdm,core,grimus1,grimus2,msgreb}, because the model
without the $\mathbf{120}$ \emph{fails}\cite{gmblm,blmdm,bert3} :
thus providing a welcome indication that it has matured
sufficiently as a scientific hypothesis so as to be  vulnerable to
falsification.

 Given these successes it is natural to ask whether -and with what
 additional constraints on its parameters- this so called New
 Minimal Susy GUT (NMSGUT)\cite{nmsgut}  is able to account for
 the longstanding discrepancy between the increasingly accurate experimental  estimates
 of the SM gauge couplings at $M_Z$ and the values predicted by
 Renormalization Group flows in the NMSGUT.  We have
 shown that the effects of threshold  corrections due to
 superheavy particles  are modest and compatible with the one
 loop architecture of MSSM coupling  unification\cite{ag2,nmsgut}.
  Moreover we found\cite{nmsgut} that  the bulk of the viable parameter
  space(according to our
   $10\%$ error allowances ) implied that the unification
 scale(and with it all other superheavy masses)
  was raised by about one order of magnitude or more thus alleviating the
  crisis regarding the too short $d=5$ proton decay lifetime estimates in Susy GUTs.

  However our analysis of the RG constraints followed a somewhat non standard
  format.   We calculated  the effects of threshold corrections on
  the values of $\sin^2 \theta_W(M_S),\alpha_G,M_{GUT}$  rather
  than the standard choices $\sin^2 \theta_W(M_Z),\alpha_G,M_{GUT}$
  (given $\alpha_3(M_Z)$) or
   $\alpha_3(M_Z),\alpha_G,M_{GUT}$ (given $ \sin^2 \theta_W(M_Z))$.
   We concluded that allowing for uncertainties $\sim 10 \%$  in
   our knowledge of the unification parameters restricted, but by
   no means eliminated, the  parameter space of this class of
   Supersymmetric GUTs.  However the
difference from the accepted format has led to some difficulty in
communicating our results to workers in the field.  Therefore in
this letter we carry out  a precision analysis in terms of the
superheavy threshold corrections to the  prediction of
$\alpha_3(M_Z),\alpha_G,M_{GUT}$.

   Our analysis is aimed   at restricting the parameter space of
 the NMSGUT based on the requirement that the parameters be such
 as to yield a lowering of the prediction of $\alpha_3(M_Z)$ from
 the two loop corrected value of $0.130$  which is
 uncomfortably larger than the central value of the
   world average experimental value $0.120 \pm
 .01$.    We  indicate  the   region of the control parameter
  space of the NMSGUT (as well as its complex parameter cousin) which yields
  threshold corrections of the right sign and magnitude to bring
  the prediction for $\alpha_3(M_Z)$ back to the observed central
  value.

\section{ Threshold corrections  }

The two loop RG flow equations for the three gauge couplings of
the standard model and MSSM  can be
integrated\cite{hall,langpolo1,langpolo2} to yield predictions for
the Grand Unification scale, the value of the gauge coupling at
unification and the value of one of the gauge couplings at the
scale $M_Z$ given the values of the other two as inputs. Since it
is $\alpha_{3}(M_Z)$ which carries the largest uncertainty ($\sim
8\%$) while $\alpha_{em}(M_Z),\sin^2\theta_W(M_Z)$ are quite
precisely known (better than $0.01\%,0.1\%$ respectively) it is
 usual\cite{langpolo1,langpolo2} to choose to predict $\alpha_{3}(M_Z)$.  Using updated
parameter values\cite{pdb} \bea M_H&=&117 GeV \qquad;\qquad
M_Z=91.1876 \pm .0021 GeV\nnu \alpha(M_Z)^{-1}&=& 127.918 \pm .018
\qquad;\qquad {\hat s }_Z^2=.23122\pm .00015 \nnu
 m^t_{pole}&=&172.7 \pm 2.9 GeV \eea

 we find from the  equations of \cite{langpolo2}
\bea \alpha_s(M_Z)- \Delta_{\alpha_s} = 0.130\pm 0.001+3.1 \times
10^{-7} GeV^{-2} \times [(m_t^{pole})^2 - (172.7 GeV)^2] +
H_{\alpha_s} \eea where
$\Delta_{\alpha_s}=\Delta_{\alpha_s}^{GUT}+\Delta_{\alpha_s}^{Susy}
$ threshold corrections.

The effect of the two loop Yukawa coupling corrections
$H_{\alpha_s}$ was estimated\cite{langpolo2} to be bounded
:$-0.003 < H_{\alpha_s}(h_t,h_b)<0$

Thus if we place credence on the central value of $0.12$ for
$\alpha_s(M_Z)$,  the effects of the low energy thresholds (
equivalently  parameterized by  an effective supersymmetry
breaking scale $M_{SUSY}$\cite{langpolo2}), GUT thresholds (and
even NRO corrections) lumped together into $\Delta_{\alpha_s}$
should correct the excessive value found. The effect of the Susy
thresholds can raise or lower the value of $\alpha_s(M_Z)$. For $
250 GeV > M_{SUSY}>20 GeV$(the actual values of superpartner
masses will be much higher since $M_{SUSY}$ is a composite
parameter formed from all the superpartner masses\cite{langpolo2})
one finds that
 $0.005 >\Delta_{\alpha_s}^{Susy} > -0.003 $. Thus it appears that
$\alpha_s(M_Z)- \Delta_{\alpha_s}^{GUT}$ could be as high as 0.135
or as low as 0.124 so that superheavy threshold corrections in the
range $-0.004>\Delta_{\alpha_s}^{GUT}>-0.015$  are indicated.

As we have described at length in earlier
papers\cite{ag2,gmblm,blmdm,nmsgut}, the superheavy thresholds are
controlled by one `fast' control parameter $ x$ and several other
`slow' parameters which are generically of order 1 or else do not
affect the thresholds significantly. Thus it is possible to scan
over  the complex $x$ plane   calculating the changes in the RG
flow predictions due to the GUT thresholds and obtain an overview
of the behaviour over the whole parameter space. By imposing
stability of the one loop analysis the viable ranges of $x$ and
even of the slow parameters can be
depicted\cite{ag2,blmdm,nmsgut}. In earlier papers we had taken
$\alpha_s(M_S)$ as input and restricted changes in the output
$\sin^2\theta_W(M_S)$ to be no more than $10\%$. We then obtained
a relatively wide range of viable parameter values. With
$\sin^2\theta_W(M_Z)$ as input and the narrow range of desired
threshold corrections as described above one gets a relatively
narrow range of permitted values of the fast parameter $x$.
However it should be remembered that the $x$ parameter is
determined by solving a cubic equation with one parameter $\xi$
which is a ratio  $\xi=\lambda m/\eta M$ of the MSGUT parameters,
so that even a fixed value of x would allow a ball of the actual
parameter values. In\cite{nmsgut} we had advocated a version of
the theory with real parameters only(with CP violation arising
spontaneously). In that case one needs to scan only over a
parameter line rather than the complex plane. In the next section
we briefly recapitulate the structure of NMSGUT to make the RG
analysis intelligible.

The mass formulae required to compute the threshold corrections
are all given in\cite{ag2,nmsgut}. We do not repeat them here but
simply furnish plots illustrating the restriction of the parameter
space due to the above described demands placed on it .

\section{Essentials of NMSGUT}
 The    NMSGUT \cite{nmsgut}  is  a
  renormalizable  globally supersymmetric $SO(10)$ GUT
 whose Higgs chiral supermultiplets  consist of AM(Adjoint Multiplet) type   totally
 antisymmetric tensors : $
{\bf{210}}(\Phi_{ijkl})$,   $
{\bf{\overline{126}}}({\bf{\Sigb}}_{ijklm}),$
 ${\bf{126}} ({\bf\Sigma}_{ijklm})(i,j=1...10)$ which   break the GUT symmetry
 to the MSSM, together with Fermion mass (FM)
 Higgs {\bf{10}} (${\bf{H}}_i$) and ${\bf{120}}$($O_{ijk}$).
  The  ${\bf{\overline{126}}}$ plays a dual or AM-FM
role since  it also enables the generation of realistic charged
fermion   and    neutrino masses and mixings (via the Type I
and/or Type II Seesaw mechanisms);  three  {\bf{16}}-plets
${\bf{\Psi}_A}(A=1,2,3)$  contain the matter  including the three
conjugate neutrinos (${\bar\nu_L^A}$).
 The   superpotential   (see\cite{abmsv,ag1,bmsv,ag2,nmsgut} for
 comprehensive details ) contains the  mass parameters
 \bea
 m: {\bf{210}}^{\bf{2}}; \qquad   M : {\bf{126\cdot{\overline {126}}}}
 ;\qquad M_H : {\bf{10}}^{\bf{2}};\qquad m_O :{\bf{120}}^{\bf{2}}
\eea

and trilinear couplings
  \bea
 \lambda &:& {\bf{210}}^{\bf{3}}; \qquad  \qquad  \eta  :
 {\bf{210\cdot 126\cdot{\overline {126}}}}
 ;\qquad  \gamma \oplus {\bar\gamma}  : {\bf{10 \cdot 210}\cdot(126 \oplus
{\overline {126}}})\nnu k &:& {\bf{ 10\cdot 120\cdot{
{210}}}};\qquad
 \rho :{\bf{120\cdot 120\cdot {  { 210}}}} \nnu \zeta
&:&{\bf{120\cdot 210\cdot{ {126}}}};\qquad \bar\zeta :
{\bf{120\cdot 210\cdot{ \overline {126}}}}
  \eea

In addition   one has two   symmetric matrices $h_{AB},f_{AB}$ of
Yukawa couplings of the the $\mathbf{10,\oot}$ Higgs multiplets to
the $\mathbf{16 .16} $ matter bilinears and one antisymmetric
matrix $g_{AB}$ for the coupling of the ${\bf{120}}$ to two
${\bf{16}}$ s. It was shown\cite{grimus1,grimus2} that with only
spontaneous CP violation, i.e with all the superpotential
parameters real, it is till possible to achieve an accurate fit of
all the fermion mass data which   furthermore
 evades the difficulties  encountered in accommodation  with
 the high scale structure of the MSGUT\cite{blmdm}
  provided\cite{core,msgreb,nmsgut,grimus1,grimus2,emerg}
one takes the $\mathbf{{10,120}}$ yukawa couplings to be much
larger than those of the $\mathbf{\oot}$ so that Type I neutrino
masses are enhanced.

The GUT scale vevs and therefore the mass spectrum are all
expressible in terms of a single complex parameter $x$ which is a
solution of the cubic equation

\be 8 x^3 - 15 x^2 + 14 x -3 = -\xi (1-x)^2 \label{cubic} \ee
where  $\xi ={{ \lambda M}\over {\eta m}} $.

Spontaneous CP violation implies that $x$ must lie\cite{nmsgut} on
one of the two complex solution branches
$x_\pm(\xi),(\xi\in(-27.917,\infty))$. Since $\lambda,\eta$ are
already counted as independent $x_+(\xi)$ counts for $M/m$.

 The other parameters besides x (or equivalently $\xi$) have
 a much  weaker   effect on the spectra and therefore the  threshold
corrections and can thus be fixed at representative values $\sim
1$ while scanning the behavior over the x-plane. Once the viable
regions are identified the limits of permissible variation in
these slow parameters can be  explored\cite{nmsgut}.

\begin{figure}[h!]
\begin{center}
\epsfxsize15cm\epsffile{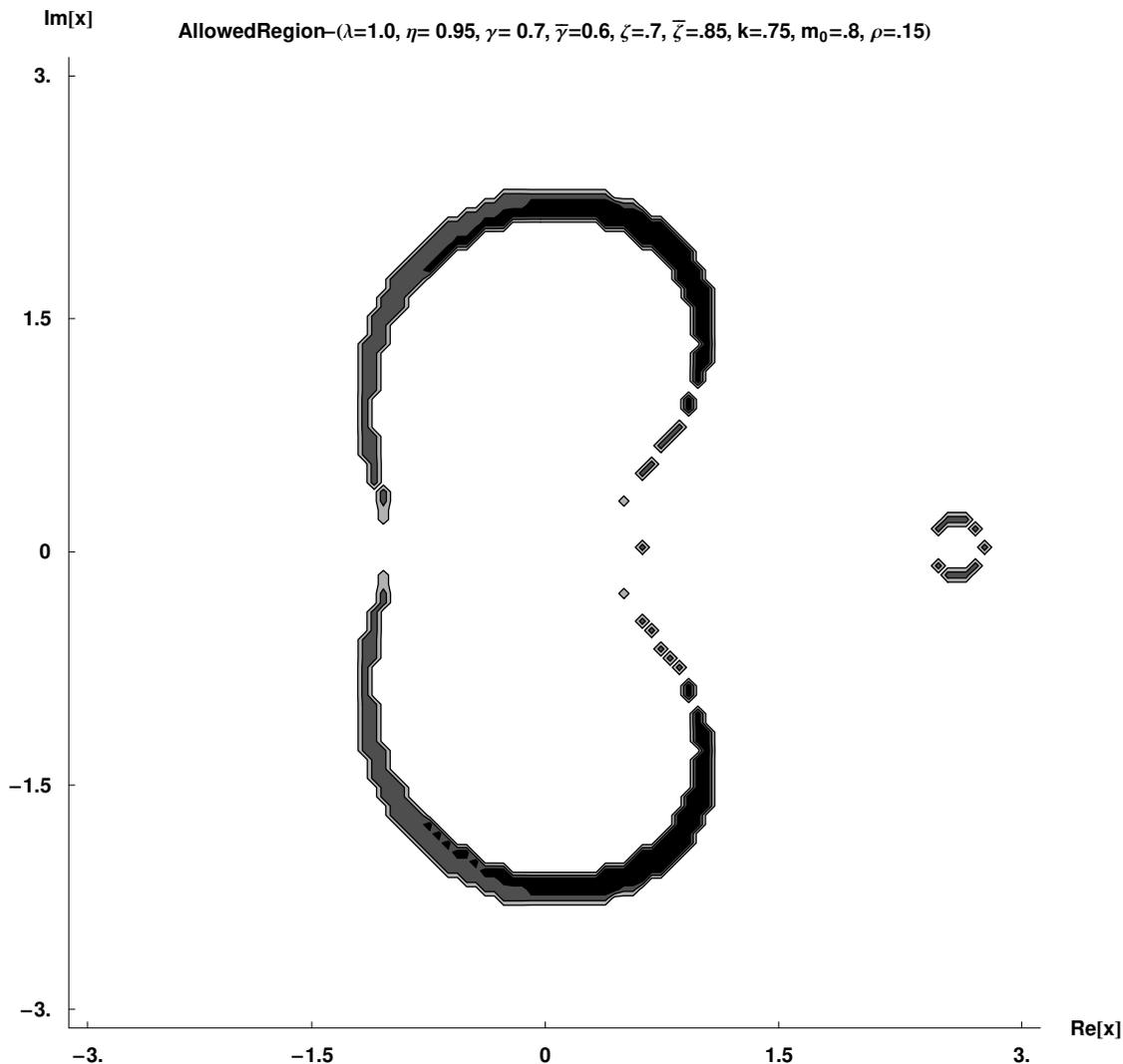} \caption{The allowed
region(other than white) for threshold corrections in the range
$-0.015<\alpha_3(M_Z)<-.005$  and for which the changes in
$\alpha_G,M_X$  are acceptable. The black region corresponds to $
1 \leq \Delta_X \leq 2$, and  dark grey to $ 0 \leq \Delta_X < 1$,
while the thin  light grey rim is for $ -1 \leq \Delta_X < 0$. }
\end{center}
\end{figure}

\begin{figure}[h!]
\begin{center}
\epsfxsize15cm\epsffile{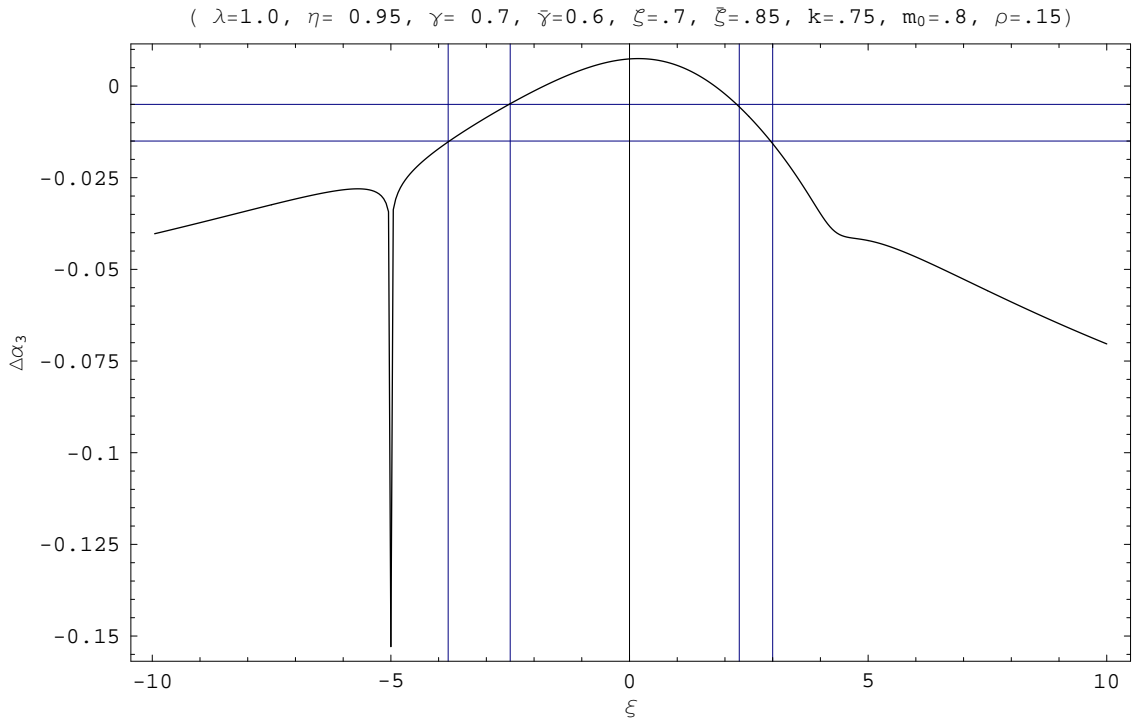} \caption{Plot of
$\Delta\alpha_3(M_Z)$ against  $\xi$ on the CP violating solution
branch $x_{+}(\xi)$ at representative values of the diagonal
parameters. Horizontal lines mark out where $\Delta \alpha_3(M_Z)$
is in the desired range $-0.015<\Delta\alpha_3(M_Z)<-.005$ and
determine the range of viable $\xi(\xi \in (-3.8,-2.5)\cup
(2.4,3))$.}
\end{center}
\end{figure}

\section{Correcting $\alpha_3(M_Z)$}

In our previous work we had taken it as self evident that, given
our demonstration that corrections to the unification parameters
 could be quite modest ($\sim 10\%$),  there would be
no problem in accommodating any particular desired values of
corrections in this range, given the number of parameters that
affected the superheavy mass spectrum. In other words fixing the
GUT superpotential parameters for the AM Higgs fields on the basis
of unification requirements alone seems a vain hope since there is
likely to remain a significant multi parameter ball compatible
with current measured values and limits. However we have recently
shown\cite{pinmsgut} that in fact proper numerical fits of the
fermion data actually determine complete  candidate NMSGUT
parameter sets. Smoking gun discoveries of supersymmetry and GUT
characteristic processes such as proton decay will, however, still
be  necessary to choose among candidate fits and thus actually fix
the NMSGUT parameters. Here we merely restrict ourselves to
displaying examples of parameter regions where the threshold
corrections are capable of lowering $\alpha_3(M_Z)$ to the current
central experimental values while maintaining the very accurately
known $\alpha_{em}(M_Z), \sin^2\theta_W(M_Z)$ at the known central
values and limiting the variation in $\alpha_G$ to $10\%$ and $ 2
>\Delta Log_{10} M_X > -1$. \textit{Note that although a negative
change of $10\%$ in $Log_{10} M_X$ would be disastrous such a
problem in fact never arises : in all cases the region allowed by
the other parameters always features a {\it{raised}} value of
$M_X$.} As we have already emphasized
previously\cite{nmsgut,emerg} this raise is in fact a welcome
mollification of the mounting tension regarding too large $d=5$
operator mediated proton decay rates. All RG formulae and mass
spectra required for our analysis have already
appeared\cite{ag1,ag2,nmsgut} and will not be repeated here.

\begin{figure}[h!]
\begin{center}
\epsfxsize15cm\epsffile{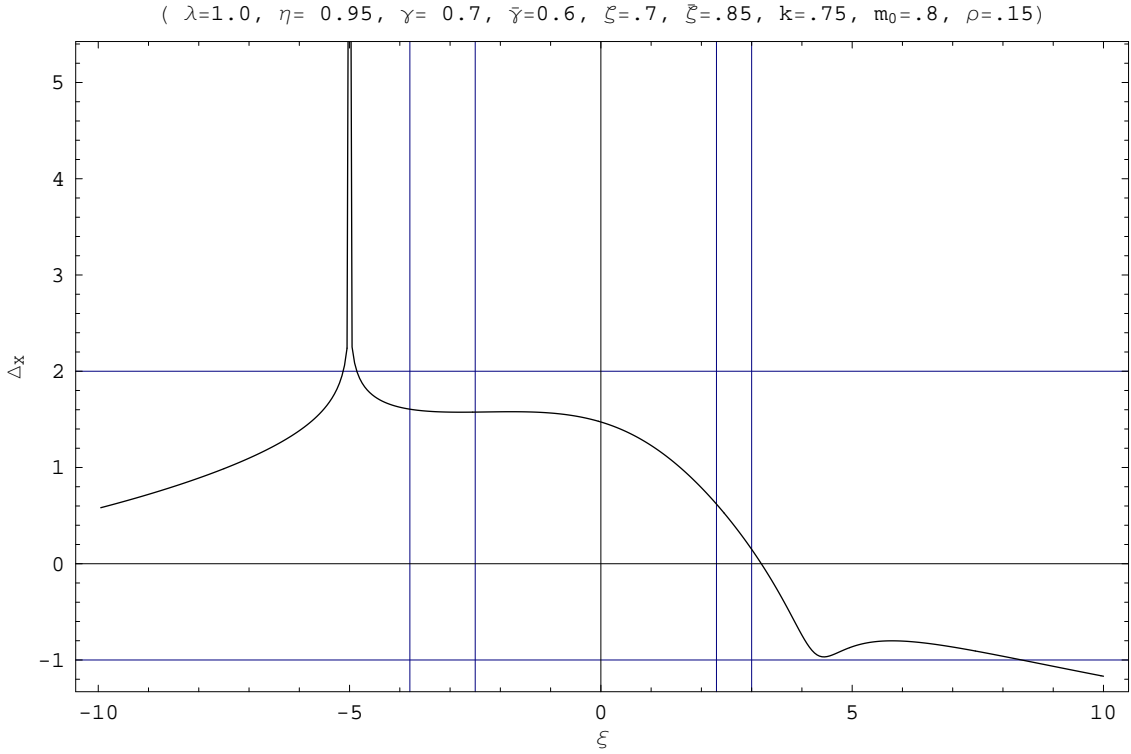} \caption{Plot of
$\Delta_X=\Delta\log_{10} M_X$ against $\xi$ on the CP violating
solution branch $x_{+}(\xi)$ at representative values of the
diagonal parameters. The regions with desired $\Delta
\alpha_s(M_Z) : -0.015<\alpha_3(M_Z)<-.005$ are marked with
vertical pairs of lines and have $M_X$ raised above the one loop
value.}
\end{center}
\end{figure}

 In Figure 1 we plot the region of the $x-$plane
which can give threshold corrections to $\alpha_3(M_Z)$of the
desired size while maintaining acceptable values for
$\alpha_G,M_X$. As already noted in \cite{nmsgut} the unification
scale is significantly raised over almost the entire allowed
range. This can be seen in Figure 1 where the  region with $
 -0.015 < \Delta \alpha_3(M_Z)< -0.005$ is stratified according to the
predicted values of $M_X$.

In\cite{nmsgut} we had identified the contour $x_+(\xi)| \xi
\in[-27.917, \infty)$ in the x plane corresponding to a complex
solution of the cubic equation  for  $x$(with positive imaginary
part) as the appropriate locus on which to examine the behaviour
of a theory in which CP violation arose spontaneously even though
initial values of all parameters were real. In practice a plot for
$\xi\in(-10,10)$ is suffcient\cite{nmsgut}. In the present context
we wish to indicate which parts of the contour allow acceptable
threshold corrections for $\alpha_3(M_Z)$ and the values of
$\alpha_G,M_X$  are within $10\%$ of their one loop values. In
Figure 2 we plot the superheavy threshold correction to
$\alpha_3(M_Z)$ versus the parameter $\xi$ on the branch $
x_+(\xi)$. In Figure 3 we plot $\Delta\log_{10}M_X$ versus $\xi$.

From these figures it is apparent that there are significant
regions of parameter space where the threshold corrections are of
the right magnitude and sign to correct the calculated 2-loop
value of $\alpha_3(M_Z)$ to its observed value. In particular we
find that the value of $M_X$ is always raised above the one loop
unification scale. Moreover this change is
accompanied\cite{nmsgut} by an parallel raising of all superheavy
masses so that the dimension 5 operator mediated proton decay
rates are reduced.\\

 \section{Discussion}
      In this letter we have illustrated how not only the general
      NMSGUT based on the Higgs system i.e ${\bf 126} \oplus {\bf
{\overline{126}}} \oplus {\bf 210}\oplus {\bf 120} \oplus {\bf
10}$, but also its more       constrained but fully realistic
version with only real   parameters, are  comfortably able to
provide the threshold  corrections of the right magnitude to
remove the discrepancy between RG corrected and experimental
$\alpha_3(M_Z)$. In fact these constraints from the RG analysis
must be confronted\cite{pinmsgut,csgrg2cum} with the determination
of GUT parameters that  arises whenever a fit of the fermion data
extrapolated to $M_X$ is  performed. The question of whether
otherwise viable fits   which will determine all parameters and
thus threshold  corrections will also keep $\Delta \alpha_3(M_Z)$
in desired range is thus  amenable to resolution in coming years.

\section{Acknowledgments}
 \vspace{ .5 true cm}
C.S.A acknowledges lively discussions with Profs. Jogesh Pati and
Stuart Raby which stimulated this analysis. The work of C.S.A was
supported by a grant No SR/S2/HEP-11/2005
 from the Department of Science and Technology of the Govt. of
 India and that of S.K.G by a University Grants Commission Junior
 Research fellowship.

\end{document}